\documentstyle[12pt]{article}
\begin{document}
\thispagestyle{empty}
\newcommand{\vsl}{v\!\!\!/}
\renewcommand{\thefootnote}{\alph{footnote}}
\begin{flushright}
{\sc BU-HEP} 93-25\\November 1993
\end{flushright}
\vspace{.1cm}
\begin{center}
{\LARGE\sc Lattice QCD and chiral Lagrangians$^{*}$}\\[2cm]
{\sc Stanley Myint and Claudio Rebbi}\\[1.5cm]
{\sc Physics Department \\[3mm]
Boston University\\[3mm]
590 Commonwealth Avenue\\[3mm]
Boston, MA 02215, USA}\\[1.5cm]
{\em * Based on a lecture given by C. Rebbi at the 1993 ICTP Summer School
 in High Energy Physics and Cosmology, to be published in the
proceedings of the School (World Scientific Publishing - Singapore -
1994).}\\[1.5cm]
{\sc Abstract}
\end{center}
\begin{quote}
After a very brief review of the formalism of lattice gauge theories
we show how one can calculate the parameters of the continuum chiral
Lagrangians proceeding through the derivation of an effective lattice
chiral Lagrangian as an intermediary step.  The derivation is done
in the strong coupling limit.  We also discuss how the derivation
could be carried out in the intermediate coupling domain by
numerical simulation techniques.
\end{quote}

\section{Introduction}

\hspace{1cm}The lattice regularization of a quantum gauge field
theory$^1$,
coupled with numerical simulation methods, provides a very powerful
tool to calculate non-perturbative observables otherwise inaccessible to
more conventional analytic techniques.  Since the introduction of lattice
gauge theory (LGT) simulations in 1979$^2$,
enormous progress has been made in the application of these techniques.
The proceedings of the international symposia on lattice gauge theories,
which have been held yearly since 1983, offer one of the best references
for recent advances in the field$^3$.
In the course of the years, LGT
calculations have increased in scope, scale and degree of sophistication,
progressing from the early estimates of the simplest observables to
reasonably accurate determinations of several quantities of
phenomenological interest.  One of the features which characterizes the
maturity of LGT simulations is that they have gone way beyond
being a (poor) substitute for the laboratory provided nature,
good only for a direct approximation of the observables of the
continuum theory.  Rather, the full power of the numerical laboratory
is being more and more exploited.  Thus, information about the dynamics
of the interactions might be obtained, for instance, from the way in which
the finite size of the lattice affects the results of a simulation.  While
the ultimate
goal is, of course, always to use the lattice regularization for deriving
the properties of the continuum, an increased interest in the dynamics of
the discretized system pays off dividends with the information it
indirectly provides about its continuum counterpart.

        In this lecture we would like to illustrate how one may derive
the parameters of the effective chiral Lagrangians$^{4,5,6}$
used for a phenomenological description
of low energy QCD interactions from first principles by lattice
techniques.  The strategy will be to exploit the fact that the lattice
regularized system also exhibits a chiral symmetry in order to perform,
directly on the lattice, the same
reduction of degrees of freedom which conceptually characterizes the
derivation of an effective Lagrangian in the continuum.
By taking the
limit of low lattice momenta in the lattice chiral Lagrangian one can then
calculate the parameters of the continuum chiral Lagrangian itself.
The derivation of the lattice chiral Lagrangian will be done here in the
strong coupling approximation.  The material which we are presenting is
taken from research which is currently in progress and a corresponding
determination of the lattice chiral Lagrangian for more realistic values
of the lattice coupling constant, which must be done numerically, has not
been carried out yet.  The strong coupling calculation, however,
illustrates even better than a numerical simulation the reduction
of degrees of freedom that leads to the lattice chiral Lagrangian
and which is at the root of the method we propose.  A discussion of how
the same reduction could be implemented numerically in the intermediate
coupling domain will be presented at the end of this paper.\\[9mm]
1.1.\hspace{4mm}{\it Lattice QCD}\hfill
\vspace{4mm}

        We begin by reviewing in an extremely concise manner the formalism
of lattice QCD.  By defining a quantum field theory on a lattice one
achieves a regularization of the ultraviolet divergences which is
non-perturbative and gauge invariant$^1$.
In general, Euclidean space-time is discretized
by the introduction of a regular lattice of points
and oriented links joining neighboring points.  In the vast
majority of applications this lattice is a hypercubical lattice,
however, for reasons that will become clear later, we will
need to work with a different lattice to carry out the strong
coupling expansion.  Therefore we will leave the detailed geometry
of the lattice unspecified for now, but we will assume that
the lattice is regular and uniform, with lattice spacing
$a$.

We will denote by $x$ the coordinates of the lattice points
and by $v$ the fundamental displacement vectors in the lattice.
Thus, for clarification, with a hypercubical lattice we would have
$x=(i_1 a, i_2 a,i_3 a,i_4 a)$ with integer $i_1, i_2,i_3,i_4 $
and $v$ would range over $ (a,0,0,0),(0,a,0,0),(0,0,a,0),(0,0,0,a)$.
The fundamental variables are the gluon\footnote{
Gauge fields are
customarily denoted by $U_{x,v}$, but eventually, following another
standard notation, we will
want to reserve the $U$ symbol for the chiral fields.  We will
therefore denote the lattice gauge fields by $G_{x,v}$.}
and the quark fields.

The gauge part of the action is constructed from the plaquette
variables, i.e. the transport factors around the elementary
closed contours of the lattice.  The plaquettes are defined
by two displacement vectors, and so we will use the notation
$G_{x,v_1 v_2}$ to denote a plaquette variable.  (For the
hypercubical lattice $G_{x,v_1 v_2}=G_{x,v_1}G_{x+v_1,v_2}
G_{x+v_2,v_1}^{\dagger}G_{x,v_2}^{\dagger}$). The gauge part of the
action\footnote{$K_1$ and $K_2$ are suitable normalization
constants, dependent on the actual shape of the lattice.}
 is then given by:

\begin{equation}
S_g= \beta K_1 \sum_{x,v_1,v_2}
{{\rm Tr} }\,(I-G_{x,v_1 v_2}),
\label{1}
\end{equation}
the coupling parameter $\beta$ being
related to the bare coupling constant by $\beta=6/g_0^2$.
The plaquette variables generalize the field strength tensor
$F_{\mu \nu}(x)$ of the continuum theory and the factors in Eq.~1
are arranged so that the lattice action reduces formally to the standard
action of the gauge field in the limit of vanishing lattice spacing
$a \to 0$ if the gauge variables are identified with continuum transport
factors through $ G_{x,v}=\exp[i g_0 A_{\mu}(x) v^{\mu}] $.

The quark fields are defined over the sites of the lattice and are given
by elements of a Grassmann algebra $\psi_x$ (color, spin and flavor
indices will frequently be left implicit). The quark matter-field
action is given by:

\begin{equation}
S_q= K_2 \sum_x \sum_v (\bar{\psi}_x\vsl
G_{x,v}\psi_{x+v}-\bar{\psi}_{x+v}\vsl G_{x,v}^{\dagger}\psi_x)
+\sum_x  \bar{\psi}_x j_x \psi_x.
\label{2}
\end{equation}

In terms of fundamental variables the (lattice regularized) quantum
expectation value of any observable $O(G,\bar{\psi},\psi)$ is given by
\begin{equation}
<O> = Z^{-1} \int DG_{x,v}\ D\bar{\psi}_x D\psi_x O(G,\bar{\psi},\psi)
\exp{ [-S_g(G) - S_q(G,\bar{\psi},\psi)]}
\label{3}
\end{equation}
with

\begin{equation}
Z =  \int DG_{x,v}\ D\bar{\psi}_x D\psi_x
\exp{ [-S_g(G) - S_q(G,\bar{\psi},\psi)]}.
\label{4}
\end{equation}

It is to be noticed that, if one considers first a lattice of finite
volume $V$, letting $V \to \infty$ only at the end of the calculations,
all of the integrals in Eqs.~3 and 4 are finite and gauge invariant.
Thus the lattice formalism provides a mathematically well defined,
non-perturbative and gauge invariant regularization of QCD. Formally,
in the limit of vanishing lattice
spacing, the above expression for the quantum averages of the observables
reduces to the continuum path integral, but, of course, the latter
is {\em per se} not well defined.  The correct continuum limit is obtained
through the process of renormalization, whereby lattice spacing $a$
and bare coupling constant $g_0$ are sent simultaneously to $0$ according
to a precise functional relationship which, because of the property
of QCD of being an asymptotically free theory, can be established by
perturbative techniques.\\[9mm]
1.2.\hspace{4mm}{\it Effective chiral Lagrangians}\hfill
\vspace{4mm}

   The most important property of QCD for low energy phenomenology is that
the theory has an approximate $SU(3) \times SU(3)$ chiral symmetry, which
would be exact for $m_u=m_d=m_s=0$, and that this symmetry is
spontaneously broken.  As a consequence, the spectrum of particle states
contains eight pseudoscalar mesons of very low mass, also called
pseudo Goldstone bosons, which would
be the Goldstone bosons of the spontaneously broken symmetry if the
masses of the lightest quarks were indeed equal to zero.

Effective chiral Lagrangians have been considered in other lectures
delivered at the ICTP summer workshops and we refer the reader to the
proceedings or to the seminal papers of Gasser and Leutwyler for more
detailed information.  We summarize here only the most basic
features of effective chiral Lagrangians and of their use for low energy
QCD.

Following Gasser and Leutwyler$^{5,6}$,
one introduces the generating functional for connected Green's functions
of QCD in the presence of scalar, pseudoscalar, vector and axial vector
sources

\begin{equation}
Z(s,p,v,a)=\int Dq D\bar{q} DG e^{i \int d^4 x {\cal L}
(q,\bar{q},G;s,p,v,a)},
\end{equation}
\begin{equation}
{\cal L}={\cal L}^0_{QCD}
-\bar{q}(x)[s(x)-i\gamma_5 p(x)]q(x)
+\bar{q}(x)\gamma^{\mu}[v_{\mu}(x)+\gamma_5 a_{\mu}(x)]q(x). \label{5}
\end{equation}

In this expression, $q,\bar{q}$ represent the lightest quark flavor triplet and
${\cal L}^0_{QCD}$ is what remains of the Lagrangian of QCD when the masses
of three lightest quarks are set to zero

\begin{equation}
{\cal L}^0_{QCD}=-\frac{1}{4 g^2} G_{\mu\nu}(x)G^{\mu\nu}(x) +
\bar{q}(x)\gamma^{\mu}[i\partial_{\mu}+ G_{\mu}(x)]q(x).  \label{6}
\end{equation}
$s(x),p(x),v_{\mu}(x)$ and $a_{\mu}(x)$ are Hermitian matrices in color
space.
The effective chiral Lagrangian is introduced to provide a description
of phenomena involving small external momenta.  Such phenomena will lead
to the excitation of the pseudo Goldstone
bosons only, and correspondingly the chiral Lagrangian will be formulated
in terms of an octet of pseudoscalar fields  $\pi^a(x)$,
collected in a unitary $3\times3$ matrix:
\begin{equation}
U(x)=\exp{[\frac{i\pi^a(x)\lambda^a}{F^0}]}, \quad U(x)U(x)^{\dagger}=1.
\end{equation}
$\lambda^a,a=1,...,8 $ are Hermitian generators of $SU(3)$, with the
normalization
\begin{equation}
{\rm Tr} (\lambda^a \lambda^b)=2\delta^{ab},
\end{equation}
$F^0$ is the pion decay
constant in the chiral limit:
\begin{eqnarray}
<0 \mid \bar{q}(x)\gamma_{\mu}\gamma_5 \frac{\lambda^a}{2}q(x)\mid \pi^b>
& = & i f_{\pi} p_{\mu} \delta^{ab}, \nonumber \\
f_{\pi}=F_0(1+O(m_q))& \approx & 93.3 MeV .
\end{eqnarray}

The chiral Lagrangian itself takes the form of a perturbative
expansion\footnote{It can be shown that
in this expansion vector and axial vector sources are of order $p$,
whereas scalar and pseudoscalar sources are of order $p^2$.}
in the external momenta $p$:
\begin{equation}
Z(s,p,v,a)=\int DU e^{i \int d^4 x {\cal L}_2(U;s,p,v,a) + {\cal
L}_4(U;s,p,v,a) + \ldots},
\end{equation}
Here ${\cal L}_2$ is
a Lagrangian involving terms of order $p^2$:
\begin{equation}
{\cal L}_2=\frac{F_0^2}{4}{\rm Tr} \{[D_{\mu}U(x)] [
D^{\mu}U(x)]^{\dagger}+\chi(x) U(x)^{\dagger}+\chi(x)^{\dagger} U(x)\}.
\label{2.10}
\end{equation}
In this expression,
\begin{equation}
D_{\mu}U(x)=\partial_{\mu}U(x)-i
r_{\mu}(x) U(x) +i U(x) l_{\mu}(x) \label{2.11}
\end{equation}
is a flavor covariant derivative and matrix $\chi(x)$ collects scalar
and pseudoscalar sources:
\begin{equation}
\chi(x)=2 B_0[s(x)+i p(x)].
\end{equation}
The constant $B_0$ is related to the quark condensate in the chiral
limit:
\begin{equation}
<0\mid \bar{u} u\mid0>_0=<0\mid \bar{d} d\mid0>_0=<0\mid \bar{s}
s\mid0>_0=-F_0^2 B_0(1+O(m_q)).
\end{equation}
Similarly, ${\cal L}_4$ is the most general Lagrangian of order $p^4$:
\begin{eqnarray}
{\cal L}_4 & = & L_1 \{{\rm Tr} [(D_{\mu}U)(D^{\mu}U)^{\dagger}]\}^2 + L_2
{\rm Tr} [(D_{\mu}U)^{\dagger} (D_{\nu}U)]{\rm Tr} [(D^{\mu}U)^{\dagger}
(D^{\nu}U)]
\nonumber \\ &  +&
L_3 {\rm Tr} [(D_{\mu}U)^{\dagger}(D^{\mu}U)(D_{\nu}U)^{\dagger}(D^{\nu}U)] +
L_4 {\rm Tr} [(D_{\mu}U)^{\dagger}(D^{\mu}U)]{\rm Tr} (\chi^{\dagger}U + \chi
U^{\dagger}) \nonumber \\ &  +&
L_5 {\rm Tr} [(D_{\mu}U)^{\dagger}(D^{\mu}U)
(\chi^{\dagger}U + \chi U^{\dagger})
]+ L_6[{\rm Tr} (\chi^{\dagger}U + \chi U^{\dagger})]^2 \nonumber \\ &  +&
L_7 [{\rm Tr} (\chi^{\dagger}U - \chi U^{\dagger})]^2
+L_8 {\rm Tr} (\chi^{\dagger} U\chi^{\dagger} U+\chi U^{\dagger}\chi
U^{\dagger}) \nonumber \\ &  -&
i L_9 {\rm Tr} [F^R_{\mu\nu}(D^{\mu}U)(D^{\nu}U)^{\dagger} +
F^L_{\mu\nu}(D^{\mu}U)^{\dagger} (D^{\nu}U)]
+L_{10} {\rm Tr} (U^{\dagger}F^R_{\mu \nu}U F^{L \mu \nu}) \nonumber \\ &  +&
H_1 {\rm Tr} (F^R_{\mu \nu}F^{R \mu \nu} + F^L_{\mu \nu}F^{L \mu \nu})
+H_2 {\rm Tr} (\chi^{\dagger} \chi).\label{20}
\end{eqnarray}
Here $F^{\mu\nu}_R,F^{\mu\nu}_L$ are field strengths for right and
left-handed fields:
\begin{eqnarray}
F^{\mu\nu}_R&=&\partial^{\mu}r^{\nu}-\partial^{\nu}r^{\mu}-
i[r^{\mu},r^{\nu}] , \nonumber \\
F^{\mu\nu}_L&=&\partial^{\mu}l^{\nu}\,-\partial^{\nu}l^{\mu}\:-
i[l^{\mu},l^{\nu}] .
\end{eqnarray}

As we see from the above, at leading order, chiral symmetry restricts the
number of terms in
the effective chiral Lagrangian to only two, so we need to know only two
constants, $F_0$ and $B_0$, in order to describe all low energy phenomena
in QCD. At next to leading order, we need ten more constants, $L_1$ -
$L_{10}$. (Constants $H_1$ and $H_2$ are of no physical significance.)
Typically, these coupling constants
are fixed by comparison of the predictions that follow from the Lagrangian
with a subset of the experimental data.  One can then derive further
predictions that can be tested against experiment or used as input
in the study of other interactions (e.g. one might need to calculate
strong corrections to electroweak matrix elements).

Of course, it would be very important to be able to calculate
the values of the coupling constants appearing in the chiral
Lagrangian from first principles, directly from the fundamental
QCD Lagrangian.  This requires that one carries out
explicitly an integration over the high energy degrees
of freedom of QCD, trading off quark and gluon fields for the
fields of the pseudo Goldstone bosons.  In the phenomenological
applications one assumes the resulting form of the Lagrangian,
but, as we have just stated above, one derives its parameters
from comparison with experiment.
What we will show in the remainder of this lecture
is that the integration over quark and gluon fields can actually be
performed explicitly in the
lattice formulation in the large $N$, strong coupling limit and that it leads
to a lattice chiral Lagrangian from which the coupling constants
of Eq.~\ref{20} can be derived.  Finally we will discuss how
it may be possible to use numerical simulations to go beyond the
strong coupling approximation and derive the chiral
Lagrangian for more physical values of the lattice coupling constant
$g_0$.

\section{Lattice chiral Lagrangian in the strong coupling approximation}

\hspace{1cm}In order to derive a lattice chiral Lagrangian one must integrate
out the high energy degrees of freedom, i.e. the quantum fluctuations of
the gauge and quark fields, re-expressing the generating functional
in terms of pseudoscalar fields conjugate to the external sources.
Thus one effectively performs a bosonization of the original theory.
For general values of the bare lattice coupling constant this can
only be done numerically, but in the strong coupling and in the
large $N$ (number of colors) limits, the bosonization can be done analytically.
The techniques for deriving an effective Lagrangian in the strong coupling
limit were laid down in pioneering papers by Kluberg-Stern, Morel,
Napoly and Petersson$^7$ and Kawamoto and Smit$^8$.  We add, however, to
the  work of these authors an ingredient that plays a crucial role
for the possibility of obtaining the terms of order $p^4$ in the chiral
Lagrangian.  As in most lattice calculations, Refs.~7 and 8 deal with
a hypercubical lattice.  While such a lattice obviously does not have
the symmetry of the continuum, as is well known all tensors of rank
up to two symmetric under the group of lattice transformations are also
symmetric under the full group of 4-dimensional rotations of Euclidean
space-time.  However, this property does not carry through to tensors of higher
order, so that, although one could expand the chiral
Lagrangian derived on
a hypercubical lattice to terms of $4^{th}$ order in lattice momentum, one
would obtain terms that cannot be identified with continuum counterparts.

Inspired by the use of lattices of higher symmetry in the theory of
lattice gases$^9$, we will formulate the theory and perform
the strong coupling expansion on a body-centered hypercubical (BCH)
lattice\footnote{ This lattice, also known as $F_4$ lattice, has been
considered in prior unrelated LGT investigations, see for
instance Refs.~10-13.  However it is the motivation for its use in the
theory of lattice gases that comes closer to the one underlying our own
work.}.
The BCH lattice can be obtained from a hypercubical (HC) lattice
in two equivalent ways.  One can define it as a HC lattice with all the
sites of odd (or even) parity removed.  Equivalently one can take as sites
of the BCH lattice all the sites of a HC lattice together with all the
centers of its elementary cells (hence the name BCH).
In the BCH lattice every site has 24 nearest neighbors.
As fundamental displacement vectors $v$
in the positive direction we will take the following:

\begin{equation}
v_{ij}^{\alpha} = \frac{\vec{e}_i+\alpha\vec{e}_j}{\sqrt{2}},
\{1\leq i <  j \leq 4, \alpha = \pm1\} .\label{18}
\end{equation}

The BCH lattice has
the largest symmetry group of all four dimensional lattices, with 1152
elements (for comparison, the symmetry group of the HC lattice has
384 elements).  As a consequence of this larger symmetry group, all
tensors of rank up to $4$ with the symmetry of the BCH lattice
are also invariant under the full 4-dimensional rotation group.
This property will be very important for our derivation.

        The use of the BCH lattice entails another advantage in the strong
coupling limit.  It is of rather technical nature and so we will mention
it only briefly.  The definition of lattice fermions encounters some
notorious problems.  A straightforward discretization of the Dirac
operator leads to the introduction of additional poles in the propagators
of the fermions at the corners of the Brillouin zone (species doubling).
Although these extra fermionic modes would in general be coupled through
the gluonic quantum fluctuations, on a hypercubical lattice there is a
fourfold degeneracy of totally decoupled
fermionic degrees of freedom.  This is akin to starting with one flavor of
fermions, but discovering that the theory actually contains four decoupled
flavors.  As a consequence, the spontaneous breaking of chiral symmetry
gives origin to 16 pseudo Goldstone bosons, even though one started with a
single flavor.  On a BCH lattice, because of the higher coordination
number of its vertices, such a separation of the fermionic fields into
four decoupled components (which is not demanded by any general theorem
on lattice fermions) does not take place and one maintains the
relationship between number of explicit flavor indices and number
of pseudo Goldstone bosons proper of the continuum theory.\footnote{It
has been argued in Ref.~11 that the use of the BCH lattice
actually worsens the problem of lattice fermions as one tries to recover
the continuum limit. This is irrelevant for our purposes, because
our motivation for its use comes solely from the need of obtaining
better symmetry properties in the strong coupling limit.  We would not
advocate using the BCH lattice in numerical simulations done in the
scaling regime, where there is good evidence for the recovery of
rotational symmetry with the ordinary HC lattice.}

Our goal is to express the generating functional of Eq.~5, formulated
over the BCH lattice (cf. Eqs.~1-4), in terms of lattice bosonic
variables $U_x=\exp{[i \pi^a_x \lambda^a/F_0]}$. We follow the
work of Refs.~7,~8, adapting it to the fact that we are on the BCH
lattice\footnote{On this lattice the values of constants in Eqs.~\ref{1}
and \ref{2} are: $K_1=K_2=1/12$.}.
Also, for the sake of brevity,
we will not include into our equations explicit
reference to the external sources $v$ and $a$ coupled to the vector and
axial vector currents.  These can be introduced via suitable chiral
transport factors defined over the links of the lattice.
We will include the results for the relevant corresponding
coefficients in ${\cal L}_4$ only in the final table.

In the strong coupling expansion, the gauge term (Eq.~\ref{1}) is
suppressed by $1/g_0^2$. Therefore, we can neglect it
to the leading order in the strong coupling expansion, and the action
reduces to Eq.~\ref{2}.
Now the gauge variables appear only in the first two terms,
moreover they appear only linearly. This means that the
product of integrals over gauge variables
factors into a product of one-link integrals:
\begin{equation}
Z=\int D\psi D\bar{\psi} e^{-\sum_x \bar{\psi}_x j_x \psi_x} \prod_{x,v}
\int dG_{x,v} e^{-W_{x,v}}.
\end{equation}
Here we have defined:
\begin{equation}
W_{x,v} = \frac{1}{12} ( \bar{\psi}_x \vsl G_{x,v}
\psi_{x+v} - \bar{\psi}_{x+v} \vsl G_{x,v}^{\dagger} \psi_x).
\end{equation}
One link integrals of this type have been calculated for the case when
the gauge field couples to a bosonic source in Ref.~14. In
Ref.~7, it was shown that this result applies equally to an
interaction of gauge fields with fermionic fields. The interested reader should
consult Ref.~7 for the details. Here we just give the
final result
for the generating functional to the leading order in the strong
coupling, large-N limit:
\begin{equation}
Z=\int D\psi D\bar{\psi} \exp{\{N \sum_{x,v} {\rm Tr}[F(\lambda(x,v))]-
\sum_x \bar{\psi}_x j_x \psi_x\}}. \label{3.18}
\end{equation}
In this equation, $\lambda(x,v)$ is a matrix in Dirac-flavor space,
defined by its
elements\footnote{Greek letters will denote indices in the direct
product of Dirac and flavor spaces, and Latin letters the color indices.
In this section we will take the number of flavors to be $n$ and
will specialize to $n=3$ only at the end.}:
\begin{equation}
\lambda(x,v)_{\alpha\beta}=\frac{1}{9 N^2} \sum_{i,j=1}^N
\sum_{\delta,\gamma,\epsilon=1}^{4 n} [\bar{\psi}_i^{\alpha}(x)
\psi_i^{\delta}(x) (\vsl )_{\gamma\delta}
\bar{\psi}_j^{\gamma}(x+v) \psi_j^{\epsilon}(x+v) (\vsl )_{\beta\epsilon}].
\end{equation}
$F(\lambda)$ is the following matrix function of $\lambda$:
\begin{equation}
F(\lambda)=1-\sqrt{1-\lambda}+\ln\frac{1+\sqrt{1-\lambda}}{2},
\end{equation}
which should be truncated for finite $N$ as a power series in
Grassmann variables.
Therefore, Eq.~\ref{3.18} is valid only in the large-N limit.

Next we proceed to convert the fermionic path integral into one over
bosonic variables as in Refs.~7,~8. In
Eq.~\ref{3.18}, $N {\rm Tr} [F(\lambda)]$ is the interaction term. It
is important to notice that it depends only on color singlet
combinations $\bar{\psi}_i^{\alpha}(x) \psi_i^{\delta}(x)$ at different
points on the lattice. Since these appear also in the source term,
$-\sum_x \bar{\psi}_x j_x \psi_x$, we can rewrite the full generating
functional as:
\begin{equation}
Z=e^{ N \sum_{x,v} {\rm Tr}[F(\lambda(x,v))] } \int D\psi D\bar{\psi}
e^{-\sum_x \bar{\psi}_x j_x \psi_x}, \label{3.22}
\end{equation}
with:
\begin{equation}
\lambda(x,v)_{\alpha\beta}=\frac{1}{9 N^2}
\sum_{\delta,\gamma,\epsilon=1}^{4 n} [ \frac{\partial}{\partial
j_x^{\alpha\delta}} (\vsl )_{\gamma\delta}
\frac{\partial}{\partial j_{x+v}^{\gamma\epsilon}}(\vsl
)_{\beta\epsilon} ]. \label{3.22p}
\end{equation}
Now we have to perform the integration over fermionic variables. This
integral splits into a product of one-site integrals:
\begin{equation}
Z_0\equiv\int D\psi D\bar{\psi} e^{-S_j}=\prod_x \int d\psi_x
d\bar{\psi}_x e^{-\bar{\psi}_x j_x \psi_x}=\prod_x z_0(j_x),
\end{equation}
\begin{equation}
z_0(j_x)\equiv \int d\psi_x d\bar{\psi}_x e^{-\bar{\psi}_x j_x \psi_x}.
\end{equation}
There are two methods of integration over the fermion fields. The first
method$^7$ is based on the use of Laplace transforms. We will
follow the method of Ref.~8, where it was
shown that $z_0(j_x)$ can be written in the following form:
\begin{equation}
z_0(j_x)=\int d{\cal M}_x e^{N {\rm Tr}(j_x {\cal M}_x)-N {\rm Tr}
(\ln {\cal M}_x)
+ const}.
\end{equation}
Here ${\cal M}_x$ is a unitary bosonic matrix in Dirac-flavor space, the
trace is over Dirac-flavor indices and the constant is irrelevant so it
will be omitted hereafter.

Using Eqs.~\ref{3.22} and \ref{3.22p} we obtain an expression for the full
generating functional:
\begin{equation}
Z=\int D{\cal M} e^{ N \{\sum_{x,v} {\rm Tr} [ F(\lambda_{x,v}) ] + \sum_x
{\rm Tr} ( j_x {\cal M}_x) - \sum_x {\rm Tr} (\ln {\cal M}_x) \} }\equiv \int
D{\cal M}
e^{S({\cal M})} ,\label{3.29}
\end{equation}
where $\lambda$ is now a bosonic matrix:
\begin{equation}
\lambda(x,v)=\frac{1}{9} \vsl {\cal M}_{x+v} \vsl
{\cal M}_x.\label{31}
\end{equation}
We look for a translationally invariant saddle point of the action
in (\ref{3.29}) of the form:
\begin{equation}
{\cal M}_{x,v}^0=v, \label{4.3}
\end{equation}
$v$ being proportional to the unit matrix in Dirac-flavor space. From
this we obtain the effective potential:
\begin{equation}
V_{eff}=-\frac{S({\cal M}_{x,v}^0)}{volume}.
\end{equation}
The saddle point for the massless case is given by:
$v(a\bar{m}=0)=\sqrt{23}/4\approx 1.20$.
We will parameterize the field ${\cal M}_x$ in a nonlinear way$^8$:
\begin{equation}
{\cal M}_x=v \exp{ [\frac{i}{F_0}(S_x+\pi_x \gamma_5 + V_x^{\nu}
\gamma_{\nu} + i A_x^{\nu} \gamma_{\nu} \gamma_5 + \frac{1}{2}
T_x^{\nu\rho}\sigma_{\nu\rho})]}.
\end{equation}
Here $S,\pi,V^{\nu},A^{\nu}$ and $T^{\nu\rho}$ are sixteen
Hermitian matrices in the flavor space. By finding the propagators of these
fields, one can show that only $\pi$ are Goldstone bosons.

In principle, we would need to integrate out the scalar, vector, axial vector
and tensor fields in order to obtain an effective action for pseudoscalar
mesons. In first approximation we will neglect these
contributions\footnote{They
 are suppressed by the masses of relevant resonances.}, and concentrate
only on the direct interactions among Goldstone bosons\footnote{In this paper
we will not consider the effects of the U(1) anomaly in any detail,
so we will restrict our attention to the traceless part of $\pi_x$.}:
\begin{equation}
{\cal M}_x=v \exp{(i \frac{\pi_x \gamma_5}{F_0})}.
\end{equation}

We will expand the action in Eq.~\ref{3.29} up to the fourth order in Taylor
series around the vacuum. This will be sufficient to extract
coefficients of effective chiral Lagrangian to $O(p^4)$ since the
deviation from the vacuum $\lambda_{x,v}-\lambda_0$ is of $O(p)$.

The action in Eq.~\ref{3.29} can be written as:
\begin{equation}
S = N \sum_{x,v} \sum_{n=1}^4\{\frac{1}{n!}\left(\frac{\partial^n
F}{\partial \lambda^n}\right)_{\lambda_0}
{\rm Tr} [(\lambda_{x,v}-\lambda_0)^n]\} + N v \sum_x {\rm Tr} (j_x {\cal
M}_x).
\label{4.11}
\end{equation}

We will evaluate the Dirac trace in a basis where
$\gamma_5={\rm diag}(1,1,-1,-1)$. By using the relation
\begin{equation}
\vsl {\cal M}_{x+v} = {\cal M}_{x+v}^{\dagger} \vsl,
\end{equation}
which follows from the definition of ${\cal M}$,
we can take the Dirac traces to obtain:
\begin{equation}
{\rm Tr} [(\lambda_{x,v}-\lambda_0)^n]=2 \lambda_0^n
 {\rm Tr} _f[(U_{x+v}^{\dagger} U_x-1)^n +
\;{\rm h.c.}\;].
\end{equation}
Here we have defined new variables which are matrices in flavor space only:
\begin{equation}
U_x\equiv\exp{(i \frac{\pi_x}{F_0})}. \label{4.16}
\end{equation}
${\rm Tr} _f$ indicates a trace over flavor indices, and from now on we will
omit
the subscript $f$.
Evaluating the second term in Eq.~\ref{4.11} is straightforward and one
obtains the expression for the action in terms of new variables:
\begin{eqnarray}
S&=& 2 N \sum_{x,v}
\sum_{n=1}^4\frac{\lambda_0^n}{n!}\left(\frac{\partial^n F}
{\partial \lambda^n}\right)_{\lambda_0}
{\rm Tr} [(U_{x+v}^{\dagger} U_x-1)^n +
\;{\rm h.c.}\;] \nonumber \\
&+& 2 N v \sum_x {\rm Tr} (\bar{m} U_x +\;{\rm h.c.}\;).\label{akcija}
\end{eqnarray}
 From now on we will denote by a subscript $v$ any
variable which under the lattice symmetry transforms in the same way as
the link $v_{ij}^{\alpha}$ on which it is defined. For example:
\begin{equation}
a_v \equiv \frac{a_i + \alpha a_j}{\sqrt{2}}, a_v^2\equiv (a_v)^2,
\;{\rm etc.}\; \label{52}
\end{equation}
Partial derivatives in the direction of the link $v_{ij}^{\alpha}$ are
defined similarly:
\begin{equation}
\partial_v a \equiv \frac{\partial_i a + \alpha \partial_j a}{\sqrt{2}},
\partial_v^2 a \equiv \partial_v (\partial_v a),
\;{\rm etc.}\;
\end{equation}
These derivatives\footnote{Here we are using
the convention that, when there is no subscript $x$, a variable (or its
derivative), is evaluated at $x$.}
 appear in the Taylor series expansion of
$U_{x+v}$ in powers of $a$:
\begin{equation}
U_{x+v}=U + a (\partial_v U) + \frac{a^2}{2}(\partial_v^2 U) +
\ldots,
\end{equation}
One can now substitute the previous Taylor series expansions into
Eq.~\ref{akcija}. Using the following results for summation of
tensors of rank two and four (which are easy to prove from definitions
in Eq.~\ref{52}) over the twelve links in the positive
direction:
\begin{equation}
\sum_v a_v b_v = 3 \sum_{\mu} a_{\mu} b^{\mu},\label{610}
\end{equation}
\begin{equation}
\sum_v a_v b_v c_v d_v = \frac{1}{2}\sum_{\mu,\nu}(a_{\mu} b^{\mu}
a_{\nu} b^{\nu} +a_{\mu} b_{\nu} a^{\mu} b^{\nu} +a_{\mu} b_{\nu}
a^{\nu} b^{\mu} ),\label{620}
\end{equation}
one obtains a manifestly
Lorentz invariant expressions for the action. We will not give the
details\footnote{The interested reader should consult our longer
paper, currently in preparation, where we give all the details as
well as a discussion on how one includes currents.}
of this rather straightforward but tedious calculation and
just give the final result in the following table.
\begin{table}[h]
\caption{Comparison of the strong coupling expansion with experimental
results from Refs.~6,~18.
The left column contains our results from the strong
coupling expansion. The right column contains the experimental values
\protect{$L_i^r(m_{\eta})$}. The numbers are in units of
\protect{$10^{-3}$}. }
\begin{center}
\begin{tabular}{|c|rc|rcl|}\hline
       &{\bf theory}&&\multicolumn{3}{c|}{{\bf experiment}} \\ \hline
$L_{1}$&0.5 $(F_{0}a)^{2}$&&0.9 &$\pm$& 0.3\\
$L_{2}$&1.1 $(F_{0}a)^{2}$&&1.7 &$\pm$& 0.7\\
$L_{3}$&10.4 $(F_{0}a)^{2}$&&-4.4 &$\pm$& 2.5\\
$L_{4}$&0 $(F_{0}a)^{2}$&&0 &$\pm$& 0.5\\
$L_{5}$&0 $(F_{0}a)^{2}$&&2.2 &$\pm$& 0.5\\
$L_{6}$&0 $(F_{0}a)^{2}$&&0 &$\pm$& 0.3\\
$L_{7}$&0.9 $(F_{0}a)^{2}$&&-0.4 &$\pm$& 0.15\\
$L_{8}$&2.6 $(F_{0}a)^{2}$&&1.1 &$\pm$& 0.3\\
$L_{9}$&-6.7 $(F_{0}a)^{2}$&&7.4 &$\pm$& 0.2\\
$L_{10}$&-6.7 $(F_{0}a)^{2}$&&-5.7 &$\pm$& 0.3\\ \hline
\end{tabular}
\end{center}
\end{table}

\section{Discussion}

\hspace{1cm}We have chosen to express the entries in Table~1. in terms of a
common factor $(F_0a)^2$ because of a point we will be making below, but
the value of this quantity is also determined by the strong coupling
expansion, from the $O(p^2)$ terms in the lattice chiral Lagrangian, and
is given by $2.1 N$.  With 3 colors the entries in the table come out
about one order of magnitude higher than the values derived from
experiment.  We take this to be an indication that the strong
coupling limit produces too tight a binding between quark and
antiquark, which in turn leads to an unacceptably high value when $F_0$
is expressed in terms of $a^{-1}$ (or, equivalently, to too large a
value for $a$).  If we allowed ourselves to set $a^{-1} \approx F_0$,
which would be a more reasonable scale for a strong coupling
calculation, then the theoretical predictions for several of the
coefficients would compare rather well with the experimental
results.  Of course we should not make too much of this agreement,
because we expect the values of the coefficients to depend on
detailed dynamical features of the interactions which may not be
well reproduced by the strong coupling approximation.  What is much
more important is the demonstration, through the strong coupling
expansion, of the main point we wanted to make, namely that one can
formulate a chiral effective Lagrangian on the lattice and that
this can be the vehicle for the derivation of the parameters of the
continuum chiral Lagrangian.

Another point that should be mentioned is that the expansion for
small lattice momenta which we have used to derive the coefficients
corresponds to using the lattice chiral Lagrangian in the tree
approximation.  We have calculated a few of the corrections that would
be induced by one loop diagrams and have found these to be small.
However, the same factor of $(F_0 a)^2$ which we discussed above, enters
also as a coefficient in the denominator of the loop diagrams, so that the
statement that the loop corrections are small should be taken with
caution. They would become larger if one could assign to $F_0 a$ the
smaller value which we advocated above.  However, even if the loop
corrections turned out to be large in a more realistic calculation,
this would only represent a technical problem and not a conceptual
difficulty for the whole approach.  The very important fact is that,
because we are working with a regularized theory, all physical
quantities are well defined and finite.  Moreover, since the lattice system
also exhibits spontaneous breaking of chiral symmetry and
Goldstone bosons
in the limit of vanishing quark mass, it will
always be possible to perform an expansion for small lattice
momenta.  The coefficients of such expansion are in any case
well defined quantities.  Thus the question is whether they can be
calculated by a perturbative expansion of the chiral effective theory,
which would be more convenient, or whether, failing such possibility,
one will have to resort to numerical techniques.  Nevertheless,
even in the latter case, one would still be dealing with a bosonic
system, and would therefore avoid the need of simulating the quantum
fluctuations of fermionic variables, which represents today the major
difficulty one faces in the implementation of QCD numerical simulations.

Finally, in order to obtain reliable values for the coefficients one
should perform the calculation in the range of values of the lattice
coupling constant ($\beta= 6/g_0^2 \approx 5.7-6$, including quark
degrees of freedom in the simulation) where one witnesses the onset of
scaling to the continuum.  Such a calculation can at present only be
done by numerical techniques.  The way we envisage it could be carried
out would be by assuming a sufficiently large set of couplings for
the lattice effective theory and fixing them through the matching of
an overcomplete set of expectation values.  This is very similar to
procedures commonly used in numerical studies of renormalization
group transformations.  The crucial point is that, because the
lattice effective theory already accounts for the long range excitations
of QCD, the matching should only require a reasonably small lattice
size, of the order of the inverse $\rho$ mass.  The two theories
(the one derived from the original QCD Lagrangian and the effective
theory) should produce exactly the same values for the observables on
any lattice size, because they are mathematically equivalent.
Working with a small lattice, we are confident, one would be able to fix
the parameters of the lattice effective theory with a good degree
of accuracy and would thus establish  a solid base for the
subsequent determination of the parameters of the continuum
chiral effective Lagrangian.\\[9mm]
{\bf Acknowledgments} This research was supported
in part under DOE grant DE-FG02-91ER40676, DOE contract
DE-AC02-89ER40509,
NSF contract PHY-9057173, and by funds from the Texas National Research
Laboratory Commission under grant RGFY91B6.
We would like to thank Richard Brower, Christophe Bruno,
Sekhar Chivukula, Andrew Cohen, Gerhard Ecker, Barry Holstein,
Dimitris Kominis, Yue Shen, Jan Smit, Akira Ukawa and Steven Weinberg for
useful and stimulating discussions.


\begin{thebibliography}{99}
\bibitem{wilson}K. G. Wilson, Phys. Rev. {\bf D10} (1974) 2445.
\bibitem{creutzetal}M. Creutz, L. Jacobs and C. Rebbi, Phys. Rev. Lett.
{\bf 42} (1979) 1390; Phys. Rev. {\bf D20} (1979) 1915.
\bibitem{proceedings}
{\it Lattice 90}, Nucl. Phys. {\bf B} (Proc. Suppl.) {\bf 20} (1991);
{\it Lattice 91}, Nucl. Phys. {\bf B} (Proc. Suppl.) {\bf 26} (1992);
{\it Lattice 92}, Nucl. Phys. {\bf B} (Proc. Suppl.) {\bf 30} (1993);
{\it Lattice 93}, Nucl. Phys. {\bf B} (Proc. Suppl.) to be published.
\bibitem{weinberg}S. Weinberg, Phys. Rev. {\bf 166} (1968) 1568.
\bibitem{gl1}J. Gasser and H. Leutwyler, Ann. Phys. {\bf 158} (1984) 142.
\bibitem{gl2}J. Gasser and H. Leutwyler, Nucl. Phys. {\bf B250} (1985) 465.
\bibitem{klus}H. Kluberg-Stern, A. Morel, O. Napoly and B. Petersson,
Nucl. Phys. {\bf B190} (1981) 504.
\bibitem{kaws}N. Kawamoto and J. Smit, Nucl. Phys. {\bf B192} (1981) 100.
\bibitem{boghosian}B. M. Boghosian, Nucl.~Phys.{\bf B}~(Proc.
Suppl.){\bf 30}~(1993)~204.
\bibitem{celmaster}W. Celmaster, Phys. Rev.~{\bf D26}~(1982)~2955;
Phys.~Rev.~{\bf D28}~(1983)~2076; Phys.~Rev.~Lett.~{\bf 52}~(1984)~403.
\bibitem{celkrausz}W. Celmaster and F. Krausz, Phys. Rev.~{\bf D28}~(1983)~
1527.
\bibitem{neuberger}H. Neuberger, Phys. Lett.~{\bf B199}~(1987)~536.
\bibitem{bhanot}G. Bhanot, K. Bitar, U. M. Heller and H. Neuberger, Nucl.
Phys.~{\bf B343} (1990) 467.
\bibitem{bregross}E. Br\'{e}zin and D. J. Gross, Phys. Lett.~{\bf B97} (1980)
120.
\bibitem{ecker}G. Ecker, J. Gasser, A. Pich and E. de Rafael, Nucl.
Phys.~{\bf B321}~(1989)~311.
\bibitem{espriu}D. Espriu, E. de Rafael and J. Taron, Nucl.~Phys.~{\bf B345}~
(1990)~22; \hfil

erratum:~Nucl.~Phys.~{\bf B355}~(1991)~278. \hfil
\bibitem{bruno}J. Bijnens, C. Bruno and E. de Rafael, Nucl.~Phys.~{\bf B390}~
(1993)~501.
\bibitem{meissner}U. G. Meissner, Bern preprint BUTP-93\slash 01.
\end{thebibliography}
\end{document}